\documentclass{aa}  
\usepackage{bm} 
\usepackage{mathrsfs}
\usepackage{graphicx}
\usepackage{txfonts}
\usepackage{xcolor}
\usepackage[colorlinks=true,citecolor=blue,linkcolor=blue,]{hyperref}
\usepackage{multirow}
\usepackage{mathrsfs}
\usepackage{lineno}
\usepackage{makecell}

\begin{document} 


\title{The impact of stellar winds and tidal locking effects on the habitability of Earth-like exoplanets around M dwarf stars}
\titlerunning{The impact of stellar winds and tidal locking effects on the habitability of exoplanets around M dwarfs}

\author{J. P. Hidalgo\inst{1}\thanks{\email{juanpablo.hidalgo@uniroma1.it}} \and D. R. G Schleicher\inst{1} \and D. P. González \inst{2}}

\institute{Dipartimento di Fisica, Sapienza, Università di Roma, Piazza le Aldo Moro 5, 00185 Roma, Italy. \and Departamento de Astronomía, Facultad Ciencias Físicas y Matemáticas, Universidad de Concepción, Av. Esteban Iturra s/n Barrio Universitario, Concepción, Chile}

   \date{Received XXX; accepted XXX}
 
  \abstract
{We present an assessment of the effects of stellar wind magnetic and mechanical components on the habitability of Earth-like exoplanets orbiting the inner and outer radii of the habitable zone (HZ) of M dwarfs. We consider stars with masses in the range of $0.09 - 0.75 M_\odot$ and planets with a surface dipolar magnetic field of 0.5 G. We estimate the size of the magnetospheres of such exoplanets using the pressure balance equation including the contribution of magnetic and ram pressures from stellar winds. We explore different scenarios, including fast and slow stellar winds, to assess the relevance of kinetic contribution. Furthermore, the effect of tidal locking and potential deviations from the Parker spiral, typically used to describe the interplanetary magnetic field, are analyzed. We show that for low mass stars ($M < 0.15 M_\odot$), the ram pressure exerted by stellar winds affects the size of the magnetosphere more than the stellar wind magnetic pressure. Interestingly, when the ram pressure is not much stronger than the magnetic pressure, typically for higher mass stars, the inclusion of ram pressure can be beneficial to the magnetosphere due to the magnetopause currents. A magnetosphere with the size of that of modern Earth is difficult to achieve with the current assumptions. However, an early Earth magnetosphere is achieved by roughly half of our hypothetical planets orbiting the outer radius of the HZ in most of the considered cases. We find that deviations from the Parker spiral can affect the results significantly, reducing the magnetosphere by $56\%$ in extreme cases. Most of the hypothetical planets are most likely (or might be) tidally locked, with the notable exception of those orbiting the outer HZ of GJ 846 and V1005 Ori. }


\maketitle

\section{Introduction}

M dwarfs are the lowest mass main-sequence stars, and the predominant stellar population in our galaxy, accounting for roughly 70-75\% of all stars \citep{Henry2006,Bochanski2010}. Due to their low luminosities, M dwarfs have their habitable zone (HZ), defined as the region around a star where a planet can maintain liquid water on its surface \citep{1959AmSci..47..397H,Kasting1993, Selsis2007}, much closer than the rest of stars, providing a good contrast to detect exoplanets in transit searches \citep{2018haex.bookE.117D}. Additionally, their low masses and the proximity of their habitable zone translate to shorter orbital periods and a larger radial velocity signal, within the current spectrograph sensitivities \citep{2014PNAS..11112641K}. 
These favorable detection conditions, combined with the abundance of M dwarfs, have led to the majority of currently known potentially habitable exoplanets being found around these stars \citep{2022planetMdwarf}.

As M dwarfs can either have thick convective envelopes or be fully convective \citep{Kippenhahn2013,Kochukhov2021}, turbulent convective dynamos are expected to be operating in their stellar interior \citep[see e.g.,][]{Donati2008b,Kapyla2021, Ortiz2023}. Furthermore, these dynamos remain active for a long fraction of their lives, generating strong magnetic fields on the order of kilogauss (kG) 
\citep{Morin2010,Lang2014,Shulyak2019}. 
As a consequence of this magnetic activity, M dwarfs have a high rate of strong flares \citep{Hilton2011,2018ApJ...867...71L,Galleta2025}. Potentially habitable planets are particularly susceptible to these phenomena due to their proximity to the host star, where the effects of stellar activity are more intense and the stellar wind is dense \citep{2007AsBio...7..167K, 2007AsBio...7..185L,2010Icar..210..539Z}. 
Strong stellar emissions could lead to progressive atmospheric erosion, ultimately resulting in the complete loss of the planetary atmosphere, threatening its habitability. Nevertheless, the extent to which such emissions affect the habitability of exoplanets orbiting M dwarfs remains a subject of debate \citep[e.g.,][]{Vida2017}.

Tidal locking may also play a significant role in determining the habitability of these planets, as estimations suggest that at such short orbital distances, habitable planets are likely to be synchronous rotators \citep{Griessmeir2005,Barnes2017}.
Tidally locked planets receive highly uneven stellar heating, which can lead to atmospheric collapse on planets with insufficiently dense atmospheres \citep{2016SSRv..205..285M}.

One potential way to protect a planetary atmosphere from stellar activity is through the presence of an intrinsic planetary magnetic field. In principle, a sufficiently large magnetosphere could shield the planetary atmosphere from stellar magnetic activity 
\citep[e.g.,][]{2007AsBio...7..185L,See2014}.  
In the super-Alfvénic regime, the extent of a planetary magnetosphere can be estimated by the balance between the pressure exerted by the stellar winds and the planetary magnetic pressure, where thermal, ram, and magnetic pressures are parts of the total contributions. 
Unfortunately, very little is known about the stellar winds of M dwarfs, primarily due to the observational challenges involved in their detection \citep{Wood2004, Vidotto2017, 2019MNRAS.482.2853J}. Previous analytic studies have used the pressure balance method to estimate the magnetospheric size of Earth-like exoplanets orbiting the HZ of M dwarfs, considering either the stellar magnetic pressure \citep{Vidotto2013} or the ram pressure of the stellar wind \citep{Griessmeir2005,Griessmeir2009}. \citet{2019winds} investigated atmospheric erosion of planets orbiting GKM-type stars, with a theoretical framework that considered the effect of both pressures. Some of their estimates differ from the results of \citet{Vidotto2013}, indicating a potential contribution from the ram pressure of stellar winds. For a complete model, it is thus important to include the combined effect of both contributions.

In this work, we assess the habitability of an Earth-like exoplanet located at the inner and outer edges of the HZ around M dwarfs, exploring the impact of stellar winds. We estimate the size of the magnetosphere of such a planet assuming different contributions to the total stellar pressure as a way to analyze which contribution is more relevant and affects the habitability predominantly. We further explore additional effects, such as tidal locking and potential deviations from the Parker spiral model of stellar winds \citep{1958ApJ...128..664P,1965SSRv....4..666P}, which is used in most of the theoretical descriptions of interplanetary magnetic fields \citep[see][for a review]{2019AnGeo..37..299L}. The stellar sample, our model, and the chosen cases are described in Section~\ref{model}. The analysis and
results of the estimations are provided in Section~\ref{results}. The discussions and conclusions of this work are presented in Sections~\ref{discussions} and \ref{conclusions}, respectively.

\section{The model \label{model}}
In this section, we describe the sample of M dwarfs considered, the framework used to describe the habitable zone, and the models applied to characterize the magnetic activity and tidal locking effects in our habitability assessment.

\renewcommand{\arraystretch}{1.1} 
\begin{table*}[h!]
\centering
\caption{Stellar sample \citep[from the Appendix of][and the references therein]{Kochukhov2021}}
\begin{tabular}{l|c|c|c|c|c|c|cc} 
\hline\hline  
GJ, Name  & SpT & $T_\mathrm{eff}$ [K] & $\langle B_\mathrm{ZDI} \rangle$ [kG] & $P_\mathrm{rot}$ [days] & $R_\star~[R_\odot]$ & $M_\star~[M_\odot]$ & $\log (L_\star/L_\odot)$  \\
\hline 
GJ 3622                 &   M6.5    &  2740  &  $0.055$  &  1.5  &  0.11  &  0.09  &  -3.21  \\
GJ 1111, DX Cnc         &   M6.5    &  2700  &  $0.09$  &  0.459  &  0.11  &  0.1  &  -3.24  \\
GJ 412B, WX UMa         &   M6.0    &  2770  &  $0.98$  &  0.78  &  0.12  &  0.1  &  -3.12  \\
GJ 65B, UV Cet          &   M6.0    &  2900  &  $1.34$  &  0.23  &  0.137  &  0.102  &  -2.92  \\
GJ 1245B                &   M5.5    &  3030  &  $0.136$  &  0.71  &  0.14  &  0.12  &  -2.83  \\
GJ 65A, BL Cet          &   M5.5    &  3000  &  $0.338$  &  0.24  &  0.156  &  0.123  &  -2.75  \\
GJ 1156, GL Vir         &   M5.0    &  3160  &  $0.086$  &  0.491  &  0.16  &  0.14  &  -2.64  \\
GJ 1289                 &   M4.5    &  3110  &  $0.275$  &  54.0  &  0.217  &  0.184  &  -2.40  \\
GJ 51, V388 Cas         &   M5.0    &  3200  &  $1.613$  &  1.06  &  0.22  &  0.22  &  -2.34  \\
GJ 490B                 &   M4.0    &  3210  &  $0.68$  &  0.54  &  0.274  &  0.23  &  -2.14  \\
GJ 896B, EQ Peg B       &   M4.5    &  3340  &  $0.45$  &  0.4  &  0.25  &  0.25  &  -2.15  \\
GJ 251                  &   M3.5    &  3270  &  $0.027$  &  90.0  &  0.3  &  0.27  &  -2.03  \\
GJ 4247, V374 Peg       &   M4.0    &  3410  &  $0.71$  &  0.45  &  0.28  &  0.28  &  -2.02  \\
GJ 285, YZ CMi          &   M4.5    &  3300  &  $0.555$  &  2.78  &  0.29  &  0.32  &  -2.05  \\
GJ 873, EV Lac          &   M3.5    &  3480  &  $0.53$  &  4.379  &  0.3  &  0.32  &  -1.92  \\
GJ 793                  &   M3.0    &  3430  &  $0.2$  &  22.0  &  0.361  &  0.37  &  -1.79  \\
GJ 896A, EQ Peg A       &   M3.5    &  3530  &  $0.48$  &  1.06  &  0.35  &  0.39  &  -1.77  \\
GJ 388, AD Leo          &   M3.5    &  3540  &  $0.25$  &  2.24  &  0.38  &  0.42  &  -1.69  \\
GJ 358                  &   M2.0    &  3560  &  $0.13$  &  25.37  &  0.446  &  0.44  &  -1.54  \\
GJ 479                  &   M2.0    &  3560  &  $0.065$  &  24.04  &  0.446  &  0.44  &  -1.54  \\
GJ 569A, CE Boo         &   M2.5    &  3570  &  $0.103$  &  14.7  &  0.43  &  0.48  &  -1.57  \\
GJ 205                  &   M1.0    &  3660  &  $0.02$  &  33.64  &  0.501  &  0.5  &  -1.39  \\
GJ278Ca, YY Gem A       &   M0.5    &  3770  &  $0.26$  &  0.81  &  0.544  &  0.54  &  -1.27  \\
GJ278Cb, YY Gem B       &   M0.5    &  3770  &  $0.205$  &  0.81  &  0.544  &  0.54  &  -1.27  \\
GJ 9520, OT Ser         &   M1.5    &  3690  &  $0.13$  &  3.4  &  0.49  &  0.55  &  -1.40  \\
GJ 49                   &   M1.5    &  3750  &  $0.027$  &  18.6  &  0.51  &  0.57  &  -1.33  \\
GJ 846                  &   M0.0    &  3850  &  $0.038$  &  10.73  &  0.588  &  0.57  &  -1.16  \\
GJ 410, DS Leo          &   M1.0    &  3770  &  $0.084$  &  14.0  &  0.52  &  0.58  &  -1.31  \\
GJ 494A, DT Vir         &   M0.5    &  3790  &  $0.147$  &  2.85  &  0.53  &  0.59  &  -1.28  \\
GJ 182, V1005 Ori       &   M0.0    &  3950  &  $0.172$  &  4.35  &  0.82  &  0.75  &  -0.83  \\
\hline
\end{tabular} 
\tablefoot{From left to right: star name, spectral type SpT, effective temperature $T_\mathrm{eff}$, large-scale magnetic field derived from ZDI analysis
of the Stokes V spectra $\langle B_\mathrm{ZDI} \rangle$, rotation period $P_\mathrm{rot}$, stellar radius $R_\star$, stellar mass $M_\star$ and stellar luminosity $L_\star$.}
\label{table1}
\end{table*}

\subsection{Stellar sample}
In this work, we analyze the data from \citet{Kochukhov2021}, which consist of a sample of 30 M dwarfs, with masses in the range of $0.09-0.75~M_\odot$ and spectral types M0-M6.5. The spectral type (SpT) and the rotation period ($P_\mathrm{rot}$) were extracted directly from their Table~3. The adopted stellar surface magnetic fields are reconstructions from Zeeman-Doppler imaging (ZDI) of Stokes V spectra \citep[e.g.][]{1989A&A...225..456S}. The ZDI method is ideal for our estimations, as it provides a reconstruction of the large-scale magnetic field, and local phenomena that can potentially lead to overestimation, such as star spots, are not considered. These data were extracted from Table~2 of \citet{Kochukhov2021}, and correspond to data published by \citet{Donati2008a}, \citet{2008MNRAS.390..567M}, 
\citet{2009ApJ...704.1721P},
\citet{Morin2010}, \citet{2016MNRAS.461.1465H}, \citet{2017ApJ...835L...4K}, \citet{2017MNRAS.472.4563M}, 
\citet{2018MNRAS.479.4836L}, and
\citet{2019ApJ...873...69K}. It should be noted, that unlike the line-broadening (Stokes I) method, the ZDI method typically no does not provide errors, as the reconstruction is model dependent and not a direct measurement.
The temperatures, radii and masses were extracted from the sources mentioned above. However, if the quantity was not provided directly, a standard value based on the spectral class was used, extracted from the extended Table 5 of  \citet{Pecaut2013} \footnote{The extension can be found \href{https://www.pas.rochester.edu/~emamajek/EEM_dwarf_UBVIJHK_colors_Teff.txt}{here}.}. Finally, surface luminosity was estimated using the Stefan–Boltzmann law. All the stellar parameters mentioned are displayed in Table~\ref{table1}.

\subsection{Habitable zone \label{sectionHZ}}
The HZ distances can be obtained using \citep{Kasting1993,Whitmire1996}
\begin{equation}
    d = \left( \frac{L_\star/L_\odot}{S_\mathrm{eff}} \right)^{1/2}~ \mathrm{AU},\label{HZ}
\end{equation}
where $S_\mathrm{eff}$ is the effective stellar flux. To estimate this quantity, we use the 1D radiative–convective
cloud-free climate mode presented by \cite{Kopparapu2013}, which provides the following parametric equation
\begin{equation}
     S_\mathrm{eff} = S_\mathrm{\odot, eff } + a T_\star + b T_\star^2 + c T_\star^3 + d T_\star^4, \label{HZ2}
\end{equation}
where $T_\star = T_\mathrm{eff} - T_\odot$, and $T_\odot = 5780$ K. For the extension of the HZ, we adopt the “runaway greenhouse” limit and the “maximum greenhouse” limit as the inner ($r_\mathrm{i}$) and outer ($r_\mathrm{o}$) boundaries, respectively.
These limits were calculated using the empirical coefficients provided in Table~1 of \cite{Kopparapu2014} and applied to Eq.~(\ref{HZ2}), assuming an Earth-like planet with a mass of $1~M_\oplus$. The model assumes atmospheres dominated by H$_2$O near the inner edge and by CO$_2$ near the outer edge, with N$_2$ acting as a background gas. At the inner edge, the runaway greenhouse limit represents the point at which the planetary oceans would completely evaporate, forming a thick steam atmosphere that triggers an irreversible greenhouse effect. At the outer edge, the maximum greenhouse limit corresponds to the maximum warming effect that CO$_2$ can provide; beyond this point, the increase in Rayleigh scattering due to high CO$_2$ levels leads to planetary cooling by increasing the overall albedo.

\subsection{Magnetopause standoff distance}
The magnetopause distance $r_\mathrm{M}$ corresponds 
to the point where the total stellar pressure $p_\star$ balances the total planetary pressure $p_\mathrm{p}$, defining the boundary between the planetary environment and the influence of the host star. At the magnetopause, the planetary pressure provides shielding against the pressure exerted by the stellar magnetic fields and stellar winds.
One of the main contributions to $p_\mathrm{p}$ arises from the planetary magnetic pressure, expressed as $B_\mathrm{p}^2/2\mu_0$, where $B_\mathrm{p}$ is the planetary magnetic field. In principle, other contributions such as the ram pressure due to atmospheric escape $p_\mathrm{p}^\mathrm{ram}$ should also be included.
However, in this study we will consider a regime where $B_\mathrm{p}^2/2\mu_0 \gg p_\mathrm{p}^\mathrm{ram}$ (see Section~\ref{discussions} for the validity of this regime).
Assuming that the planet orbits at a distance $r_\mathrm{orb}$ from the star, we find
\begin{equation}
    p_\star(r_\mathrm{orb}) = p_\mathrm{p}(r_\mathrm{M}) \approx \frac{B_\mathrm{p}(r_\mathrm{M})^2}{2\mu_0}, \label{eq3}
\end{equation}
where $\mu_0$ is the magnetic permeability of vacuum. As very little is known about the magnetic fields of rocky exoplanets \citep{2024arXiv240415429B}, we will assume a dipolar magnetic field, based on Earth's large-scale magnetic field, where most of its magnetic energy is concentrated in the dipolar mode \citep[e.g.][]{2011PEPI..187..157C, Ganushkina2018}, i.e.
\begin{equation}
    B_\mathrm{p}(r) = \frac{1}{2} B_\mathrm{p,0} \left( \frac{r_\mathrm{p}}{r}\right)^3, \label{dipole-planet}
\end{equation}
where $B_\mathrm{p,0}$ is the magnetic field at the pole, assumed to be 1 G throughout this work, and $r_\mathrm{p}$ is the planetary radius. Inserting Eq.~(\ref{dipole-planet}) in Eq.~(\ref{eq3}), and solving for the radius, we have
\begin{equation}
    \frac{r_\mathrm{M}}{r_\mathrm{p}}(r_\mathrm{orb}) = \sqrt[6]{\frac{B_\mathrm{p,0}^2}{8\mu_0 p_\star (r_\mathrm{orb})}} . \label{relevant}
\end{equation}
As the  magnetic pressure is the main planetary contributor to the pressure balance, Eq.~(\ref{relevant}) can be considered as a proxy to estimate the size of the upstream magnetosphere, which is the relevant side concerning the protection against stellar winds and other forms of stellar activity. 

\subsubsection{Stellar magnetic pressure}
As a first approximation, we consider only the pressure exerted by the stellar wind magnetic field, as in the analysis by \citet{Vidotto2013}. This implies $p_\star (r_\mathrm{orb}) = B_\star (r_\mathrm{orb})^2 / 2\mu_0$, where $B_\star$ is the stellar magnetic field. Under this assumption, Eq.~(\ref{relevant}) becomes
\begin{equation}
    \frac{r_\mathrm{M}}{r_\mathrm{p}}(r_\mathrm{orb}) = \sqrt[3]{\frac{B_\mathrm{p,0}}{2 B_\star(r_\mathrm{orb})}}, \label{case1}
\end{equation}
which is equivalent to Eq.~(4) of \cite{Vidotto2013}.

\renewcommand{\arraystretch}{1.1} 
\begin{table*}[t!]
\centering
\caption{Summary of our estimations}
\begin{tabular}{l|c|c|cc|cc}
\hline\hline
GJ, Name & 
\makecell{$[r_\mathrm{i}, r_\mathrm{o}]$ [AU]} & 
\multicolumn{5}{c}{$r_M(r_\mathrm{i},r_\mathrm{o})/r_\mathrm{p}$} 
\\
\cline{3-7}
& & Case 1 & 
\multicolumn{2}{c|}{Case 2} & 
\multicolumn{2}{c}{Case 3} 
\\
\cline{4-7}
& & & slow wind & fast wind & slow wind & fast wind \\
\hline
GJ 3622                 &  $[0.026 , 0.052]$  &  $[3.876 , 6.178]$  &  $[1.526 , 1.927]$  &  $[1.271 , 1.604]$  &  $[1.526 , 1.927]$  &  $[1.271 , 1.604]$ \\
GJ 1111, DX Cnc         &  $[0.025 , 0.051]$  &  $[3.227 , 5.149]$  &  $[1.222 , 1.544]$  &  $[1.018 , 1.286]$  &  $[1.222 , 1.544]$  &  $[1.018 , 1.286]$ \\
GJ 412B, WX UMa         &  $[0.029 , 0.058]$  &  $[1.505 , 2.397]$  &  $[1.408 , 1.777]$  &  $[1.173 , 1.480]$  &  $[1.381 , 1.768]$  &  $[1.165 , 1.477]$ \\
GJ 65B, UV Cet          &  $[0.036 , 0.072]$  &  $[1.440 , 2.285]$  &  $[1.130 , 1.423]$  &  $[0.941 , 1.185]$  &  $[1.122 , 1.421]$  &  $[0.939 , 1.184]$ \\
GJ 1245B                &  $[0.040 , 0.080]$  &  $[3.271 , 5.169]$  &  $[1.616 , 2.032]$  &  $[1.346 , 1.692]$  &  $[1.616 , 2.032]$  &  $[1.346 , 1.692]$ \\
GJ 65A, BL Cet          &  $[0.044 , 0.087]$  &  $[2.383 , 3.770]$  &  $[1.289 , 1.621]$  &  $[1.073 , 1.350]$  &  $[1.288 , 1.621]$  &  $[1.073 , 1.350]$ \\
GJ 1156, GL Vir         &  $[0.050 , 0.098]$  &  $[4.027 , 6.338]$  &  $[1.666 , 2.090]$  &  $[1.387 , 1.741]$  &  $[1.666 , 2.090]$  &  $[1.387 , 1.741]$ \\
GJ 1289                 &  $[0.066 , 0.130]$  &  $[2.677 , 4.219]$  &  $[5.461 , 6.857]$  &  $[4.547 , 5.709]$  &  $[3.501 , 5.352]$  &  $[3.426 , 5.029]$ \\
GJ 51, V388 Cas         &  $[0.070 , 0.139]$  &  $[1.541 , 2.422]$  &  $[2.538 , 3.182]$  &  $[2.113 , 2.650]$  &  $[1.960 , 2.846]$  &  $[1.848 , 2.530]$ \\
GJ 490B                 &  $[0.088 , 0.174]$  &  $[2.063 , 3.242]$  &  $[2.278 , 2.856]$  &  $[1.897 , 2.378]$  &  $[2.171 , 2.817]$  &  $[1.864 , 2.367]$ \\
GJ 896B, EQ Peg B       &  $[0.087 , 0.170]$  &  $[2.494 , 3.903]$  &  $[2.259 , 2.826]$  &  $[1.881 , 2.353]$  &  $[2.223 , 2.814]$  &  $[1.871 , 2.350]$ \\
GJ 251                  &  $[0.100 , 0.196]$  &  $[6.197 , 9.719]$  &  $[7.796 , 9.763]$  &  $[6.492 , 8.130]$  &  $[7.111 , 9.483]$  &  $[6.258 , 8.047]$ \\
GJ 4247, V374 Peg       &  $[0.101 , 0.198]$  &  $[2.201 , 3.437]$  &  $[2.505 , 3.130]$  &  $[2.086 , 2.606]$  &  $[2.367 , 3.078]$  &  $[2.042 , 2.591]$ \\
GJ 285, YZ CMi          &  $[0.098 , 0.193]$  &  $[2.289 , 3.587]$  &  $[3.923 , 4.911]$  &  $[3.267 , 4.089]$  &  $[2.935 , 4.299]$  &  $[2.792 , 3.863]$ \\
GJ 873, EV Lac          &  $[0.113 , 0.220]$  &  $[2.492 , 3.882]$  &  $[4.505 , 5.623]$  &  $[3.751 , 4.682]$  &  $[3.221 , 4.760]$  &  $[3.096 , 4.342]$ \\
GJ 793                  &  $[0.132 , 0.258]$  &  $[3.384 , 5.280]$  &  $[6.946 , 8.677]$  &  $[5.784 , 7.225]$  &  $[4.428 , 6.714]$  &  $[4.336 , 6.326]$ \\
GJ 896A, EQ Peg A       &  $[0.136 , 0.263]$  &  $[2.624 , 4.081]$  &  $[3.696 , 4.609]$  &  $[3.077 , 3.838]$  &  $[3.183 , 4.365]$  &  $[2.882 , 3.761]$ \\
GJ 388, AD Leo          &  $[0.148 , 0.287]$  &  $[3.274 , 5.089]$  &  $[4.559 , 5.684]$  &  $[3.796 , 4.733]$  &  $[3.952 , 5.400]$  &  $[3.567 , 4.645]$ \\
GJ 358                  &  $[0.176 , 0.340]$  &  $[4.101 , 6.371]$  &  $[8.120 , 10.121]$  &  $[6.761 , 8.427]$  &  $[5.352 , 8.038]$  &  $[5.218 , 7.511]$ \\
GJ 479                  &  $[0.176 , 0.340]$  &  $[5.167 , 8.027]$  &  $[8.023 , 10.001]$  &  $[6.681 , 8.327]$  &  $[6.479 , 9.159]$  &  $[6.020 , 8.043]$ \\
GJ 569A, CE Boo         &  $[0.170 , 0.330]$  &  $[4.448 , 6.908]$  &  $[7.490 , 9.334]$  &  $[6.237 , 7.773]$  &  $[5.684 , 8.229]$  &  $[5.386 , 7.369]$ \\
GJ 205                  &  $[0.208 , 0.401]$  &  $[7.934 , 12.286]$  &  $[9.445 , 11.753]$  &  $[7.864 , 9.786]$  &  $[8.800 , 11.495]$  &  $[7.654 , 9.711]$ \\
GJ278Ca, YY Gem A       &  $[0.240 , 0.459]$  &  $[3.506 , 5.409]$  &  $[4.402 , 5.468]$  &  $[3.666 , 4.553]$  &  $[4.019 , 5.306]$  &  $[3.535 , 4.505]$ \\
GJ278Cb, YY Gem B       &  $[0.240 , 0.459]$  &  $[3.795 , 5.855]$  &  $[4.402 , 5.468]$  &  $[3.666 , 4.553]$  &  $[4.137 , 5.363]$  &  $[3.581 , 4.523]$ \\
GJ 9520, OT Ser         &  $[0.207 , 0.398]$  &  $[4.296 , 6.646]$  &  $[5.967 , 7.422]$  &  $[4.969 , 6.180]$  &  $[5.181 , 7.051]$  &  $[4.673 , 6.064]$ \\
GJ 49                   &  $[0.222 , 0.426]$  &  $[7.408 , 11.437]$  &  $[8.969 , 11.144]$  &  $[7.468 , 9.279]$  &  $[8.306 , 10.873]$  &  $[7.250 , 9.200]$ \\
GJ 846                  &  $[0.270 , 0.515]$  &  $[6.838 , 10.522]$  &  $[8.171 , 10.136]$  &  $[6.804 , 8.440]$  &  $[7.603 , 9.905]$  &  $[6.619 , 8.373]$ \\
GJ 410, DS Leo          &  $[0.229 , 0.439]$  &  $[5.109 , 7.883]$  &  $[8.535 , 10.601]$  &  $[7.107 , 8.827]$  &  $[6.519 , 9.369]$  &  $[6.165 , 8.380]$ \\
GJ 494A, DT Vir         &  $[0.236 , 0.451]$  &  $[4.269 , 6.582]$  &  $[6.077 , 7.545]$  &  $[5.060 , 6.283]$  &  $[5.200 , 7.116]$  &  $[4.721 , 6.147]$ \\
GJ 182, V1005 Ori       &  $[0.395 , 0.750]$  &  $[4.272 , 6.552]$  &  $[7.968 , 9.868]$  &  $[6.635 , 8.217]$  &  $[5.543 , 8.142]$  &  $[5.358 , 7.505]$ \\
\hline
\end{tabular} 
\tablefoot{From left to right: star name, inner ($r_\mathrm{i}$) and outer ($r_\mathrm{o}$) radii of the Habitable Zone (HZ), and the magnetopause standoff distance of an Earth-like planet located at the inner and outer edges of the HZ ($r_\mathrm{orb} =r_\mathrm{i}, r_\mathrm{o}$). The estimations are presented for three cases corresponding to Eqs.~(\ref{case1}), (\ref{case2}),  and (\ref{case3}). The last two cases include stellar wind contributions with slow wind ($v_\mathrm{esc}$) and fast wind ($3v_\mathrm{esc}$).}
\label{table2}
\end{table*}

To estimate $B_\star(r_\mathrm{orb})$, we assumed a reference radius at the stellar corona after which the only non-vanishing magnetic field component is the radial component. This simulates the stretching of magnetic field lines by the stellar wind, and it is based on the potential field source surface extrapolation technique \citep[PFSS, e.g.,][]{1969SoPh....9..131A,Jardine2002,Lang2012,Vidotto2013}. In the analytical models of \cite{1958ApJ...128..664P,1965SSRv....4..666P}, the radial component decays with the square of the radius, i.e.
\begin{equation}
    B_\star(r) = B_{r,\star}(r) = B_\mathrm{ref} \left( \frac{R_\mathrm{ref}}{r} \right)^2 \label{Bstar}
\end{equation}
where $R_\mathrm{ref}$ and $B_\mathrm{ref} = B_\star(R_\mathrm{ref})$ are the radius and the stellar magnetic field at the mentioned reference point. To extrapolate the surface magnetic field into the reference point, we assume a dipolar behavior for the stellar magnetic field, similar as \citet{Wanderley2024}. Therefore, $B_\mathrm{ref}$ can be estimated as 
\begin{equation}
    B_\mathrm{ref} \approx \langle B_\mathrm{ZDI} \rangle \left( \frac{R_\star}{R_\mathrm{ref}} \right)^3. \label{BSS}
\end{equation}
In this work, we assume $R_\mathrm{ref} = 2.5 R_\star$, following \citet{Vidotto2013}, and also taking into account that $R_\mathrm{ref} < r_\mathrm{orb}$. Inserting Eq. (\ref{BSS}) in Eq. (\ref{Bstar}), it yields
\begin{equation}
    B_\star (r) = 0.4 \langle B_\mathrm{ZDI} \rangle \left( \frac{R_\star}{r} \right)^2. \label{Bstarfinal}
\end{equation}

\subsubsection{Ram pressure \label{ram}}
The magnetopause can also be strongly influenced by stellar wind ram pressure \citep{Chapman1931}.
In this case, we assume that such pressure is the main contribution to the total stellar wind pressure, i.e. $p_\star = p_\star^\mathrm{ram}$, where
\begin{equation}
    p_\star^\mathrm{ram}(r) = \rho_\mathrm{w}(r) v_\mathrm{w}^2
\end{equation}
is the stellar wind ram pressure, with $v_\mathrm{w}$ being the terminal velocity of the stellar wind, and
\begin{equation}
    \rho_\mathrm{w}(r) = \frac{\dot{M}_\star}{4\pi v_\mathrm{w} r^2}\label{densitywind}
\end{equation}
is the stellar wind density, where $\dot{M}_\star$ is the stellar mass-loss rate. It is important to note that both quantities are challenging to constrain, as stellar winds are notoriously difficult to detect observationally, and numerical models predict a wide range of values that can differ by several orders of magnitude \citep[e.g.,][]{Cranmer2011, Cohen2014, Garraffo2016, Sakaue2021}. To estimate the mass-loss rate, we adopt the scaling relations proposed by \cite{Johnstone2015a, Johnstone2015b}: 
\begin{equation}
    \dot{M}_\star = \dot{M}_\odot R_\star^2 \Omega_\star^{1.33} M_\star^{-3.36},  \label{mass-loss}
\end{equation}
where $\dot{M}_\odot = 1.4 \cdot 10^{-14} ~M_\odot \, \mathrm{yr^{-1}}$ is the solar wind mass loss rate and 
$\Omega_\star$ is the stellar rotation rate. For the wind velocity, a common assumption is to scale this quantity with the escape velocity, i.e. $v_\mathrm{w} = \alpha v_\mathrm{esc}$, where $\alpha$ is a constant, typically assumed to be $1$ \citep[e.g.,][]{Cranmer2011, Suzuki2018}, and $v_\mathrm{esc} = \sqrt{2 G M_\star / R_\star}$. This scaling is based on the fact that the terminal wind velocities of the Sun are comparable to its surface escape velocity. Here, we assume $\alpha = 1$ for the slow wind ($v_\mathrm{w}^\mathrm{slow}$), and $\alpha = 3$ for the fast wind ($v_\mathrm{w}^\mathrm{fast}$). These assumptions give $v_\mathrm{w}^\mathrm{slow} = 638$ km/s, $v_\mathrm{w}^\mathrm{fast} = 1914$ km/s and $\dot{M}_\star = 5.71 \cdot 10^{-14}~M_\odot\,\mathrm{yr^{-1}}$ for EV Lac. For comparison, \cite{Sakaue2021} estimated velocities of around 900 km/s, and \cite{Cohen2014} a mass loss rate of $3\cdot 10^{-14}~M_\odot\,\mathrm{yr^{-1}}$. Observations based on the Ly$\alpha$ lines estimate $\dot{M}_\star \approx 2 \cdot 10^{-14} ~M_\odot \, \mathrm{yr^{-1}}$ for EV Lac \citep{Wood2004,Wood2021}.

The dynamical pressure of the stellar wind compresses the planetary magnetic field on the upstream side, inducing currents flowing across the magnetopause \citep{Chapman1931,Chapman1941,2015AnGeo..33..965L}. These currents generate a secondary magnetic field $B_\mathrm{mc}$, oriented in opposition to the stellar magnetic field, which enhances the shielding effect and contributes to the protection of the planetary atmosphere. Thus, Eq.~(\ref{eq3}) becomes
\begin{equation}
p_\star(r_\mathrm{orb}) = \frac{1}{2\mu_0} (B_\mathrm{p}(r_\mathrm{M}) + B_\mathrm{mc}(r_\mathrm{M}))^2.    
\end{equation}
In the original formulation of \cite{Chapman1931} with a planar magnetopause, the magnetic field produced by the so-called Chapman-Ferraro (or magnetopause) currents, has roughly the same strength as the planetary dipolar magnetic field. However, in order to consider magnetopause models with different geometries, it is convenient to introduce a form factor $f_0$, where \citep{Griessmeier2004}
\begin{equation}
    B_\mathrm{p} + B_\mathrm{mc} \equiv 2 f_0 B_\mathrm{p} \label{Bmc}.
\end{equation}
In this context, $f_0 \approx 1$ corresponds to the planar magnetopause of \cite{Chapman1931} and $f_0 = 1.5$ to a spherical magnetosphere. More realistic geometries are closer to an ellipsoid, due to the compression of the upstream side \citep[e.g.][]{1989P&SS...37.1037T}. In this work we adopt $f_0 = 1.16$, as given in \cite{volland1995handbook}, to consider deviations from a spherical magnetosphere \citep[see also][]{Griessmeier2004,Gunell2018}. Finally, the magnetopause standoff distance is then given by
\begin{equation}
    \frac{r_\mathrm{M}}{r_\mathrm{p}}(r_\mathrm{orb}) = \sqrt[6]{\frac{ f_0^2 B_\mathrm{p,0}^2}{2\mu_0 \rho_\mathrm{w} (r_\mathrm{orb}) v_\mathrm{w}^2}} \label{case2}
\end{equation}
when considering only the ram pressure of the stellar winds.

\subsubsection{Total contribution \label{total-contribution}}
Finally, the total stellar pressure corresponds to the combined effect of the magnetic pressure, ram pressure, and thermal pressure of electrons and protons, which yields $p_\star (r_\mathrm{orb}) = B_\star(r_\mathrm{orb})^2/2\mu_0 + p_\star^\mathrm{ram}(r_\mathrm{orb}) + p_\star^\mathrm{th}(r_\mathrm{orb})$. Neglecting the thermal pressure $p_\star^\mathrm{th}$, and using the formulation of Section~\ref{ram} for the ram pressure, we obtain
\begin{equation}
    \frac{r_\mathrm{M}}{r_\mathrm{p}}(r_\mathrm{orb}) = \sqrt[6]{\frac{ f_0^2 B_\mathrm{p,0}^2}{B_\star(r_\mathrm{orb})^2 + 2\mu_0\rho_\mathrm{w}(r_\mathrm{orb}) v_\mathrm{w}^2 }}.\label{case3}
\end{equation}

\begin{figure*}[h!]
    \centering
    \includegraphics[width=\hsize]{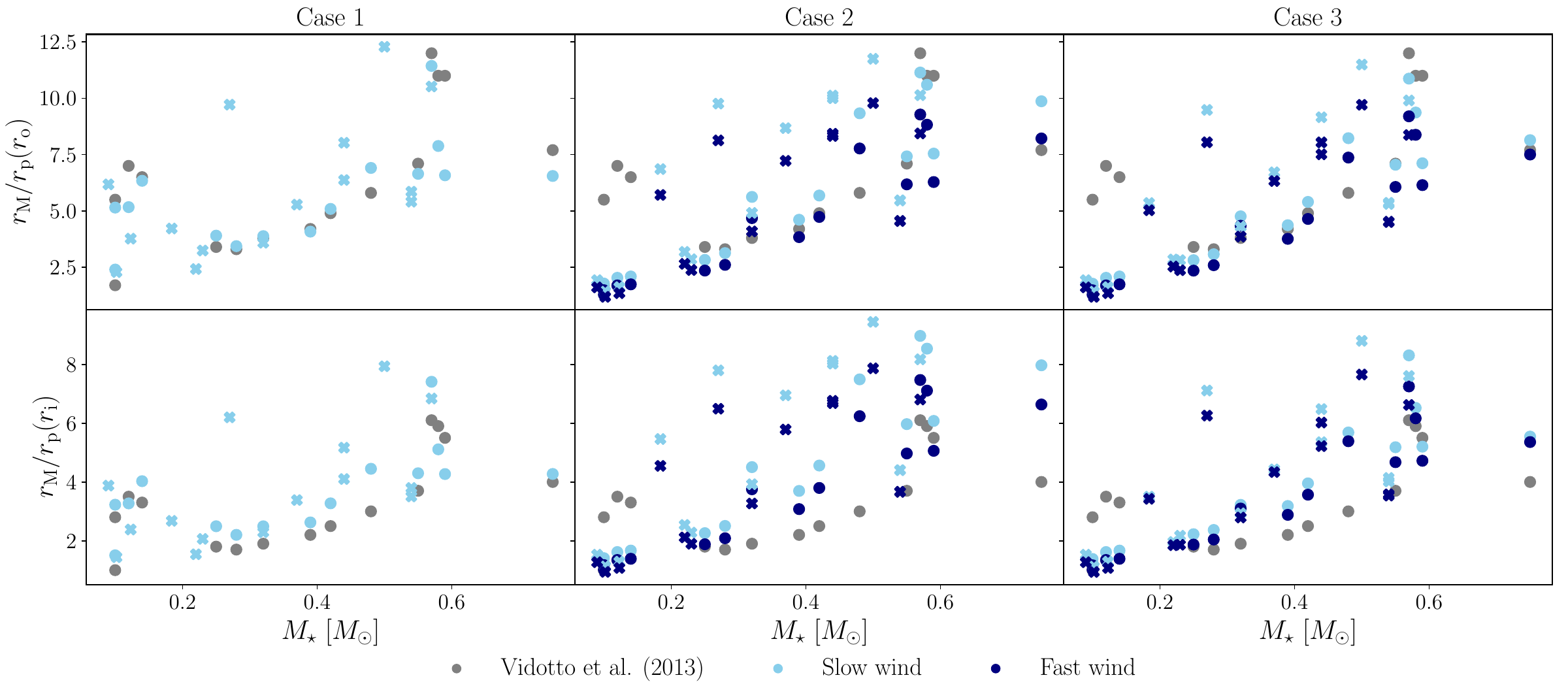}
    \caption{The magnetopause standoff distance of an Earth-like planet ($B_\mathrm{p,0} = 1$ G) orbiting the 30 M dwarfs from our sample, estimated in Case 1 (\textit{left panels}), Case 2 (\textit{middle panels}) and Case 3 (\textit{right panels}), for slow ($v_\mathrm{w} = v_\mathrm{esc}$) and fast ($v_\mathrm{w} = 3v_\mathrm{esc}$) winds. The results of \cite{Vidotto2013} were also added for comparison. In the \textit{upper (bottom) panels} the planet is orbiting the outer (inner) radius of the HZ. The dot marker indicates that the star is in the sample of \cite{Vidotto2013}, and the x marker that it was not included in their work.}
    \label{fig:plot1}
\end{figure*}

\subsection{Tidal locking \label{TL0}}
To estimate the timescale required for a planet to be tidally locked, we use the following expression \citep{Guillot1996, Griessmeir2005}
\begin{equation}
\tau_\mathrm{lock} \approx Q \left( \frac{r_\mathrm{p}^3}{G M_\mathrm{p}} \right) (\omega_\mathrm{i} - \omega_\mathrm{f}) \left( \frac{M_\mathrm{p}}{M_\star} \right)^2 \left( \frac{r_\mathrm{orb}}{r_\mathrm{p}} \right)^6, \label{TL1}
\end{equation}
where $Q$ is the tidal dissipation factor, $G$ is the gravitational constant, $M_\mathrm{p}$ is the mass of the planet, $\omega_\mathrm{i}$ and $\omega_\mathrm{f}$ are the initial and final angular velocities, respectively. As $\omega_\mathrm{f}$ is expected to match the synchronous rotation rate, it can be obtained by Kepler’s law
\begin{equation}
    \omega_\mathrm{f} = \sqrt{\frac{M_\star G}{r_\mathrm{orb}^3}}. \label{omegaf}
\end{equation}
Assuming Earth-like parameters for the hypothetical planet, it yields $r_\mathrm{p}= R_\oplus = 6378000~\mathrm{m}$, $M_\mathrm{p} = M_\oplus = 5.972 \cdot 10^{24}~\mathrm{kg}$ and $\omega_\mathrm{i} = \omega_\oplus = 7.2722 \cdot 10^{-5}~\mathrm{rad/s}$. For the Earth's $Q$ parameter, the current estimate is $Q = 12 \pm 2$ \citep{Williams1978,1984plin.book.....H}. However, this value creates inconsistencies between predictions and observations, e.g., it estimates that the Moon has been in orbit for 1-2 Gyrs \citep[see Section 2.4 of][for a discussion]{Barnes2017}. One potential way to solve this issue, is to assume an historical average for $Q$ instead of just the current value. The modern value of Earth says very little about its past history, e.g., the numerical tidal model of \cite{Green2017} predicts that during the Cenozoic and Late Cretaceous, the tidal dissipation rates were far below the present levels, which should be reflected in an increase of $Q$. This is probably a consequence of ocean configuration and continental drift, i.e. the  arrangement of continents and shallow seas of modern Earth make tidal dissipation very efficient \citep{Kasting1993, 2009ESRv...97...59C, Green2017}. To account for this possible issue, we adopted two different values of the $Q$ parameter of Earth, $Q=12$, that is the current estimation, and $Q=100$, which takes into account the potential historical evolution of it. Additionally, $Q=100$ is similar to other rocky planets, e.g. Mars has a tidal dissipation factor of $93 \pm 8.4$ \citep{2022JGRE..12707291P}, so the decision of $Q=100$ can also be interpreted as a more general description of a terrestrial exoplanet, instead of the unusually low value of modern Earth.

\section{Results \label{results}}

We estimate the magnetopause standoff distances of an Earth-like planet, orbiting the inner $r_\mathrm{i}$ and outer $r_\mathrm{o}$ edges of the HZ of the 30 M dwarfs from our data sample. We consider three cases. Case 1: only the stellar magnetic pressure is taken into account. We neglect Chapman-Ferraro currents. Eq.~(\ref{case1}) is used. Case 2: only the stellar wind ram pressure is considered. Chapman-Ferraro currents are included. Eq.~(\ref{case2}) is used. Case 3: the stellar magnetic and ram pressures are considered. Chapman-Ferraro currents are included. Eq.~(\ref{case3}) is used.

The estimations of the HZ and the three different cases for the magnetopause distance are listed in Table~\ref{table2}. In Figure~\ref{fig:plot1} we compare our results for Case 1 with the results of \cite{Vidotto2013} for their sample of 15 M dwarfs for the purpose of validation, and to show our results also for the additional stars in our sample. We further compare the results for Cases 1, 2 and 3 and consider the possibility of both a slow wind and a fast wind. In Case 1 (left panels), our results are mostly in agreement with \citet{Vidotto2013}. In the cases of V374 Peg, EV Lac, EQ Peg A and AD Leo, the magnetopause distance at the outer radius of the HZ overlaps with their results (upper left panel), where the best case is EV Lac with a difference of $0.082$ in the ratio of $r_\mathrm{M}/r_\mathrm{p}$. However, the agreement is somewhat reduced at the inner radius of the HZ. This is most likely a consequence of the different formula used to estimate the extent of the HZ. \cite{Vidotto2013} used the “early Mars” and “recent Venus” criteria from \citet{Selsis2007}, and we chose the “runaway greenhouse” and “maximum greenhouse” criteria from \cite{Kopparapu2014} (see Section~\ref{sectionHZ}). The extent of the HZ using the last criteria is typically smaller, being 75\% of the extent obtained with the former. While the values of $r_\mathrm{o}$ between both models are comparable, the difference in the estimated $r_\mathrm{i}$ is typically larger \citep[see Table~1 of][column 11]{Vidotto2013}, producing the mentioned discrepancy. 
On the other hand, in cases like DS Leo and DT Vir their estimation at the outer radius $r_\mathrm{o}$ is significantly higher, with a difference of $3.117$ and $4.418$, respectively. This does not seem to be related to the definitions of the HZ, nor the slightly different $\langle B_\mathrm{ZDI} \rangle$ chosen. One possible explanation, is that DS Leo and DT Vir have significant non-axisymmetric field components \citep{Vidotto2014}. As we are extrapolating the magnetic fields into the stellar corona assuming an axisymmetric dipolar configuration, it is probable that the estimated values at $B_\mathrm{ref}$ are larger than those estimated by \cite{Vidotto2013} using the PFSS method, reducing the size of $r_\mathrm{M}/r_\mathrm{p}$. Our results for GJ 49 are in agreement with \cite{Vidotto2013}, as both estimations in the outer HZ are close to the modern Earth value, which is considered to be $r_\mathrm{M}/r_\mathrm{p} = 11.7$. In the case of DS Leo and DT Vir, our results are closer to the early Earth magnetospheric size, $r_\mathrm{M}/r_\mathrm{p} = 5$ \citep{Tarduno2010}. It is worth mentioning that outside the sample of \citet{Vidotto2013}, we find three stars whose magnetopause distances at the outer HZ are closer to the present Earth value, which are GJ 251, GJ 846, and GJ 205, with $r_\mathrm{M}/r_\mathrm{p} =9.719$, $10.522$ and $12.286$, respectively. We further note that at the inner boundary of the HZ, all the planets have a magnetospheric size smaller than that of modern Earth, where the largest is $r_\mathrm{M}/r_\mathrm{p}=7.934$ for GJ 205.

In the middle and right panels of Figure~\ref{fig:plot1}, it is visible that in Cases 2 and 3, the discrepancy between our results and those of \citet{Vidotto2013} is typically larger. For example, in the low mass stars DX Cnc, GJ 1245 B, GJ 1156, the difference between the results for a planet orbiting the outer HZ in Case 2, assuming a fast wind, is 4.214, 5.308 and 4.759, respectively. In the inner HZ, the differences are smaller, being 1.782, 2.154 and 1.913, respectively. In some stars the agreement gets better, for example, in OT Ser, the difference between Vidotto's value and our estimation for Case 1 at $r_\mathrm{o}$ is 0.454, while in Case 3 with the slow wind it is 0.049. Outside the sample of Vidotto, the largest magnetospheres in Case 2 for planets orbiting $r_\mathrm{o}$ are GJ 358, GJ 846, GJ 205, with 10.121, 10.136 and 11.753 for the slow wind, respectively. In the cases with a fast wind, none of the magnetospheres from our sample reach a ratio of 10, and the highest values for both $r_\mathrm{i}$ and $r_\mathrm{o}$ are from GJ 205. We note in general that the results depend both on the definition of the HZ as well as the treatment of the magnetic effects, with degeneracies between the two.

\begin{figure*}[h!]
    \centering
    \includegraphics[scale=0.58]{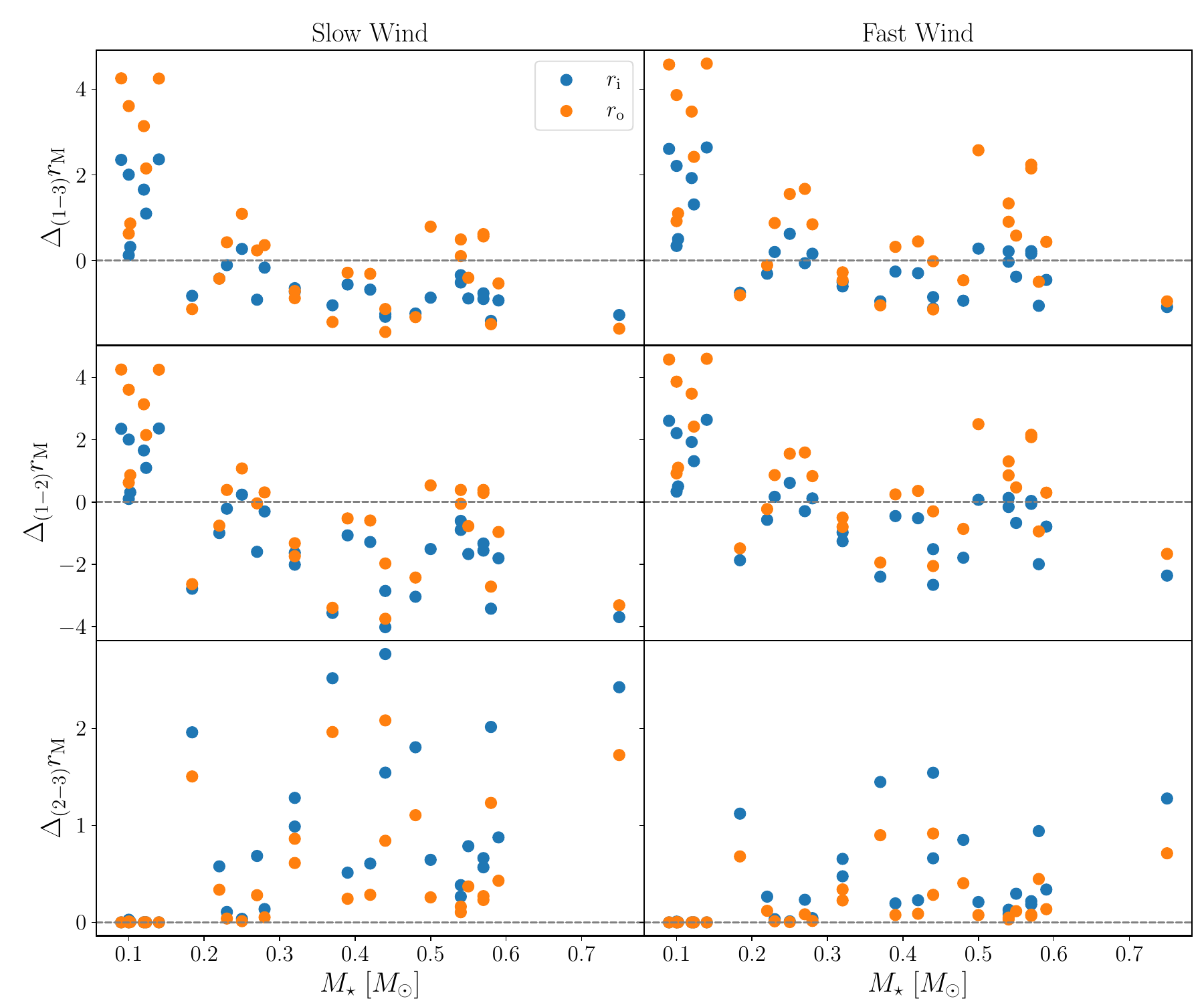}
    \caption{Differences between the three considered cases, $\Delta_\mathrm{(1-3)} r_\mathrm{M}(r_\mathrm{orb})$, $\Delta_\mathrm{(1-2)} r_\mathrm{M}(r_\mathrm{orb})$ and $\Delta_\mathrm{(2-3)} r_\mathrm{M}(r_\mathrm{orb})$ following Eqs~(\ref{delta1})-(\ref{delta3}). The blue (orange) dots represent planets orbiting the inner (outer) radii of the HZ. The results for the slow wind are given on the panels on the left, the results for the fast wind on the right.}
    \label{fig:plot2}
\end{figure*}

To quantify how much the results change depending on the considered contributions of the stellar wind pressure, we define the following variables:
\begin{align}
    \Delta_\mathrm{(1-3)} r_\mathrm{M}(r_\mathrm{orb}) &= r_M^\mathrm{case\,1}(r_\mathrm{orb})/r_\mathrm{p} - r_M^\mathrm{case\,3} \label{delta1}(r_\mathrm{orb})/r_\mathrm{p}, \\
    \Delta_\mathrm{(1-2)} r_\mathrm{M}(r_\mathrm{orb}) &= r_M^\mathrm{case\,1}(r_\mathrm{orb})/r_\mathrm{p} - r_M^\mathrm{case\,2} \label{delta2}(r_\mathrm{orb})/r_\mathrm{p}, \\
    \Delta_\mathrm{(2-3)} r_\mathrm{M}(r_\mathrm{orb}) &= r_M^\mathrm{case\,2}(r_\mathrm{orb})/r_\mathrm{p} - r_M^\mathrm{case\,3}(r_\mathrm{orb})/r_\mathrm{p}, \label{delta3}
\end{align}
where the superscript “case x” represents the chosen case for the calculation of the magnetopause standoff distance. These three variables are displayed in Fig.~\ref{fig:plot2}. In low mass stars ($M < 0.15M_\odot$), $\Delta_\mathrm{(1-3)} r_\mathrm{M}(r_\mathrm{i},r_\mathrm{o})$ and $\Delta_\mathrm{(1-2)} r_\mathrm{M}(r_\mathrm{i},r_\mathrm{o})$ obtain their maximum values, which indicates that in this mass range, our models predict that the ram pressure of the stellar wind affects the magnetosphere of the planet more than the stellar magnetic pressure, i.e. $p_\star^\mathrm{ram} \gg B_\star^2/2\mu_0$. This can also be seen in the lower panels, where $\Delta_\mathrm{(2-3)} r_\mathrm{M}(r_\mathrm{i},r_\mathrm{o})$ is essentially zero for these stars, showing that the stellar wind magnetic field has a negligible effect. This is a consequence of their high mass loss rates, which is achieved by their low masses ($\dot{M}_\star \propto M_\star^{-3.36}$). For more massive stars, most of the values of $\Delta_\mathrm{(1-3)} r_\mathrm{M}(r_\mathrm{i})$ are negative in case of the slow wind. This result is counterintuitive, as the combination of both elements is expected to decrease the size of the magnetosphere compared to Case 1, e.g., \citet{Vidotto2013} estimate that the inclusion of stellar ram pressure should decrease the radii by a factor of 2.5. However, this increase in $r_\mathrm{M}/r_\mathrm{p}$ is a direct consequence of the Chapman-Ferraro currents that arise from the interaction of the stellar wind plasma and the planetary magnetic field \citep{Ganushkina2018}. In our calculations, these currents enhance the magnetic pressure of the planet by a factor of $ \sim 5.38$, contributing to the shielding of the atmosphere. In the case of low mass stars, this additional shielding does not overcome the high ram pressure of the stellar winds. However, for stars with higher masses ($M > 0.15 M_\odot$), this results for typically $\Delta_\mathrm{(1-3)} r_\mathrm{M}(r_\mathrm{i}), \Delta_\mathrm{(1-2)} r_\mathrm{M}(r_\mathrm{i}) < 0 $ in the slow wind. In the cases with fast wind the ram pressure increases, and therefore, some cases become positive, especially in the outer HZ.

The magnetopause currents are included in Cases 2 and 3. Therefore, the values of $\Delta_\mathrm{(2-3)} r_\mathrm{M}(r_\mathrm{i},r_\mathrm{o})$ are expected to be positive. This is visible in the bottom panels of Fig.~\ref{fig:plot2}. As mentioned, in low mass stars $\Delta_\mathrm{(2-3)} r_\mathrm{M}(r_\mathrm{i},r_\mathrm{o})$ is close to zero. For the other stars, the inclusion of stellar magnetic field affects $r_\mathrm{M}/r_\mathrm{p}$ either significantly, like in GL Vir, or weakly, like in EQ Peg B, so that the tendency can vary significantly.

\subsection{Deviations from the Parker model \label{altdecay}}

\begin{figure*}[t!]
    \centering
     \includegraphics[width=\hsize]{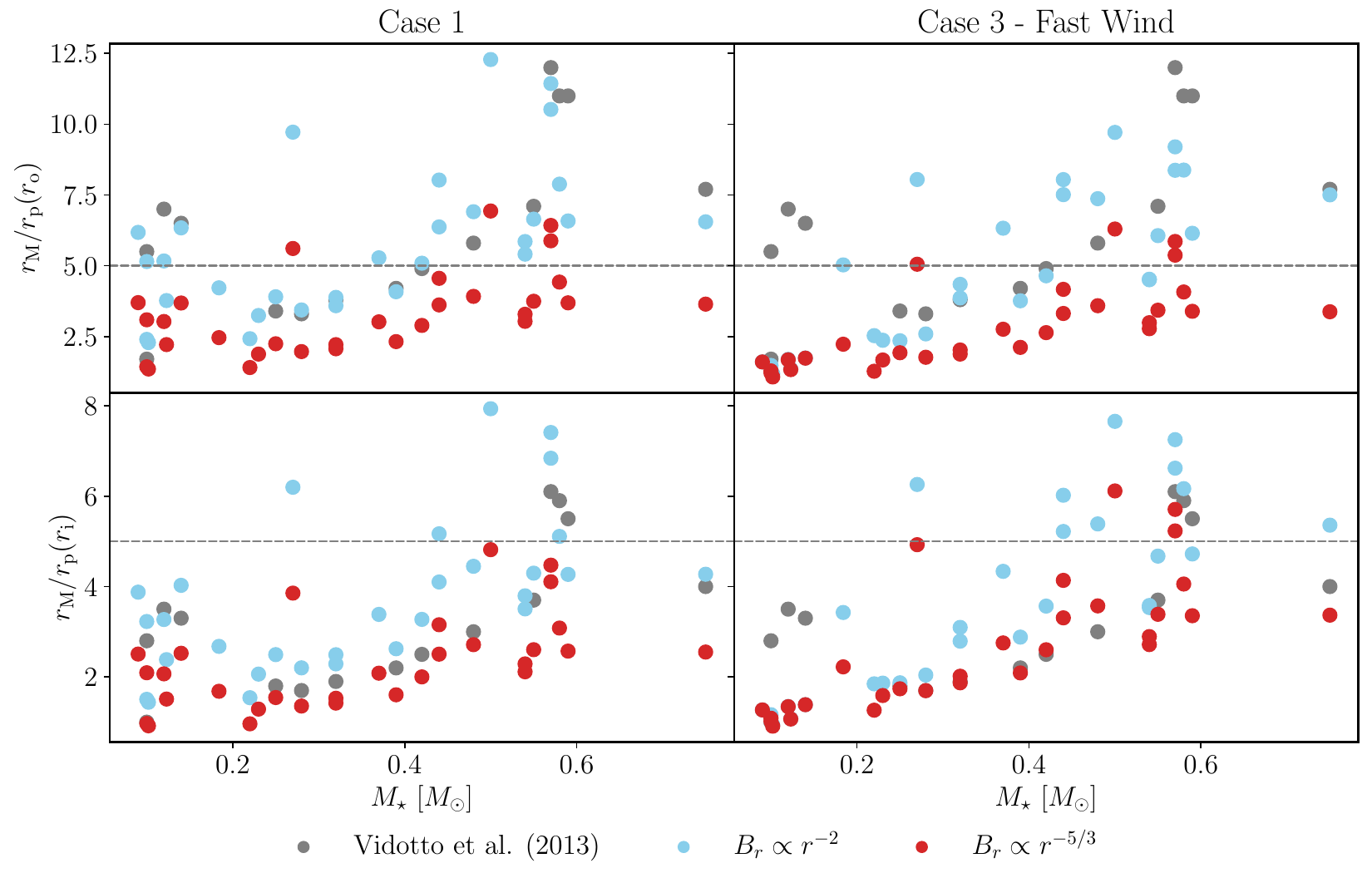}
    \caption{The magnetopause standoff distance of a planet orbiting our stellar sample. The theoretical decay of the Parker model (blue) and that measured in the solar system (red) are used in the estimations for Case 1 (\textit{left panels}) and for Case 3, fast wind (\textit{right panels}). In the upper (bottom) panels the planet is orbiting the outer (inner) radius of the HZ. The gray dashed line represents Earth's early magnetosphere. The results of \cite{Vidotto2013} were added for comparison.}
    \label{fig:plot3}
\end{figure*}

In the models of \citet{1958ApJ...128..664P, 1965SSRv....4..666P}, the assumption that the radial magnetic field decays following $B_r \propto r^{-2}$ is directly based on the assumption of magnetic flux conservation. However, in reality, this assumption is not always valid \citep{2017ApJ...848...70L}. For example, in-situ measurements from spacecrafts in the solar system have reported deviations from the Parker model, often attributed to complex phenomena like turbulence, magnetic clouds or coronal mass ejections (CMEs) \citep{2010JGRA..115.9101B, 2016SoPh..291..265W, 2024A&A...690A.233K}. This leads to a measured magnetic flux that exceeds the value predicted by theory \citep{2013ARep...57..844K}. In M dwarfs, this is highly speculative, as no direct measurements have been done. 
\cite{2020MNRAS.494.1297M} performed simulations of the stellar winds of M dwarfs, reporting that in the high-$\beta$ regime ($\beta > 1$), their models are mostly in agreement with the Parker spiral. However, in the low-$\beta$ regime ($\beta < 1$), the disagreement between models increases significantly. As M dwarfs typically have very strong magnetic fields, and their HZ is very close to them (see column 2 of Table~\ref{table2}), a low-$\beta$ regime with relevant deviations could be expected. Another relevant factor could be CMEs. Assuming 50 CMEs per day, \citet{2016ApJ...826..195K} estimated that an exoplanet around an M dwarf will be impacted 0.5–5 times per day, which is 2-20 times the mean value at Earth during solar maxima. 
However, the excess flux produced by CMEs in the Sun does not seem to be enough to explain the observed magnetic flux excess \citep{2015ApJ...809L..24W, 2017ApJ...848...70L}, and the occurrence of CMEs in M dwarfs is still under debate. In any case, to compensate the potential effects of a deviation from the Parker spiral, we use the scaling reported by \citet{2012ApJ...761...82K}, obtained from multi-spacecraft observations from 0.29 AU to 5 AU, given by
\begin{equation}
    B_r(r) = B_0 \left( \frac{r_0}{r} \right)^{5/3}. \label{obs}
\end{equation}
Assuming $r_0 = R_\mathrm{ref}=2.5 R_\star$, and $B_0 = B_\mathrm{ref}$, Eq.~(\ref{Bstarfinal}) becomes
\begin{equation}
    B_\star (r) = 0.4 \langle B_\mathrm{ZDI} \rangle \left( \frac{R_\star}{r} \right)^{5/3}. \label{BstarfinalOBS}
\end{equation}

In Figure~\ref{fig:plot3} the estimations of the magnetopause distance using the measured decay in the solar system are shown, for Cases 1 and 3. It is clearly visible that the assumed deviation from the Parker spiral produces significant differences in the final output. In Case 1, now the highest value, that comes from a planet orbiting GJ 205, is 4.816 at the inner radius $r_\mathrm{i}$ and 6.933 at the outer radius $r_\mathrm{o}$, which significantly reduces $r_\mathrm{M}/r_\mathrm{p}$ by a difference of 3.118 and 5.353, respectively, compared to our original case. Typically, in this case, the magnetopause distance gets reduced by $30-40 \%$ at the $r_\mathrm{i}$, and by $40-45 \%$ at $r_\mathrm{o}$, which can compromise the habitability of the hypothetical planets. For example, a planet located at $r_\mathrm{o}$ from most of the stars with $M_\star \geq 0.37 M_\odot $ would not achieve $r_\mathrm{M}/r_\mathrm{p} \geq 5$ anymore.


In Case 3, planets orbiting low mass stars ($M_\star < 0.15 M_\odot$) are not affected significantly by this alternative decay, reducing their magnetosphere by $0.04-3 \%$, mainly because $p_\star^\mathrm{ram} \gg B_\star^2/2\mu_0$, same as in our original case. In the rest of stars, there is no clear tendency, and the magnetospheres are reduced by $14-37 \%$ at inner radius $r_\mathrm{i}$ and by $18-56 \%$ at outer radius $r_\mathrm{o}$. Furthermore, the smallest standoff distance under these assumptions comes from UV Cet, with a ratio of 0.911. This is also the smallest $r_\mathrm{M}/r_\mathrm{p}$ in this entire work. We note that these estimations are somewhat optimistic, mainly because deviations from the Parker spiral could increase other parameters of the wind, like e.g, velocity, and mass loss rate, by orders of magnitudes, decreasing the sizes of the magnetospheres more severely than shown in Fig.~\ref{fig:plot3}.

\subsection{The effect of tidal locking \label{TL}}

\begin{figure*}[t!]
    \centering
    \includegraphics[width=\hsize]{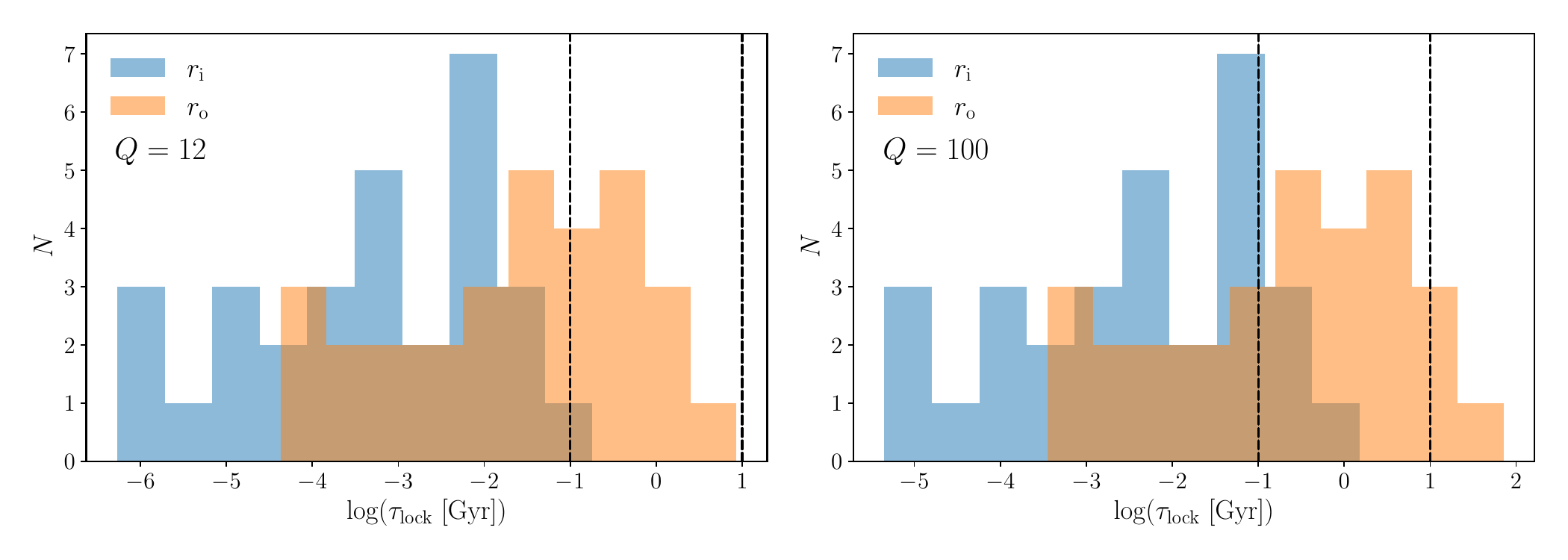}
    \caption{The timescale for tidal locking estimated for an Earth-like planet orbiting the inner and outer radii of the HZ around our stellar sample, following Eq.~(\ref{TL1}). Due to the uncertainty of the tidal dissipation parameter $Q$, we use the current value (\textit{left panel}), and an estimation based on the historical average (\textit{right panel}). The vertical dashed lines represent the different considered regimes.}
    \label{fig:Tlock}
\end{figure*}

\begin{figure*}[t!]
    \centering
    \includegraphics[width=\hsize]{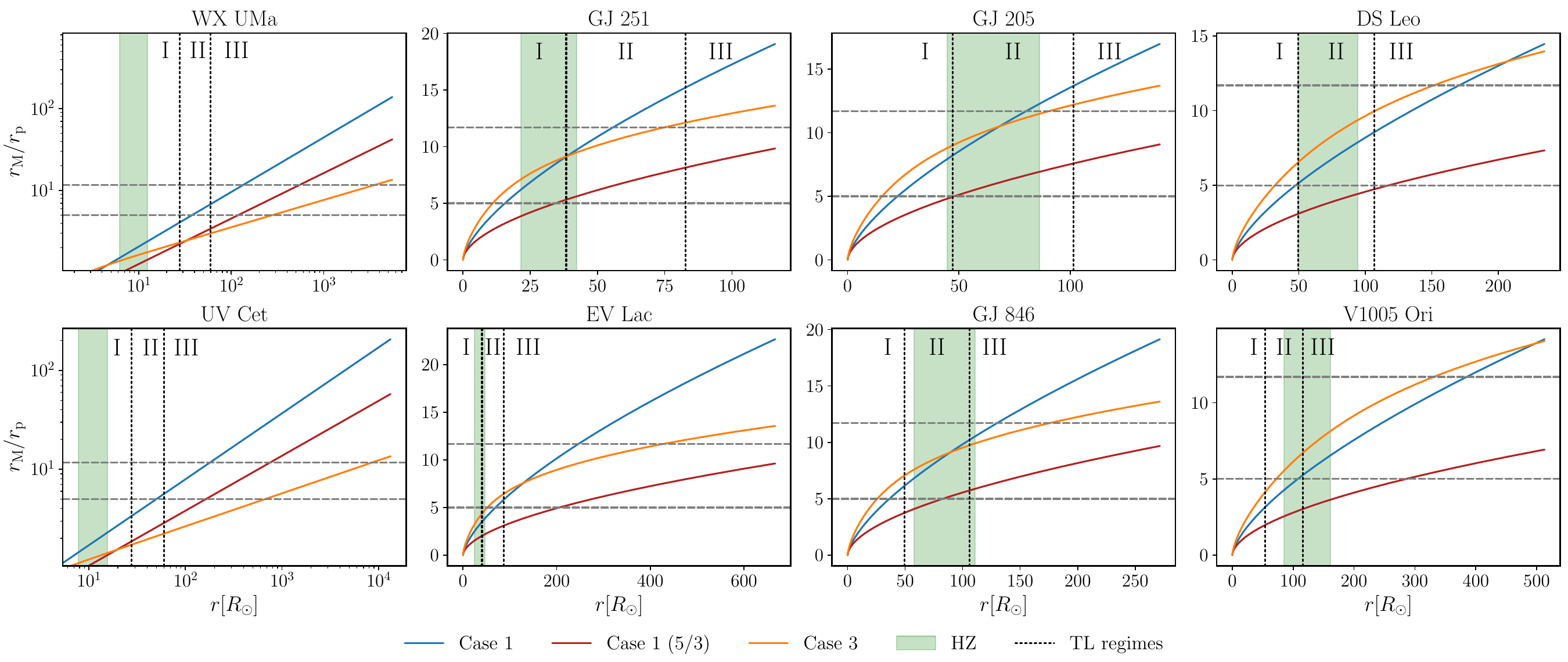}
    \caption{Magnetopause distance as a function of the stellar radius of representative stars from our sample. The blue and orange lines represent Case 1 (only stellar magnetic pressure is considered) and Case 3 (stellar magnetic and ram pressures are considered) with slow wind, respectively. The red line is Case 1 with the alternative magnetic field decay $B_r \propto r^{-5/3}$. The solid green area is the HZ of the star. The gray horizontal dashed lines indicate the size of the magnetosphere of early ($5 r_\mathrm{p}$) and modern Earth ($11.7 r_\mathrm{p}$). The black vertical dotted lines represent the different tidal locking regimes assuming $Q=100$ (see Section~\ref{TL}).}

    \label{fig:plot5}
\end{figure*}

Following Eq.~(\ref{TL1}) and using the planetary parameters discussed in Section~\ref{TL0}, we estimated the tidal locking timescales for an Earth-like planet orbiting the habitable zone of the M dwarfs from our sample. The results for the two chosen $Q$ parameters are shown in Figure~\ref{fig:Tlock}. To quantify if the planet is tidally locked or not, the regimes of \citet{Griessmeir2005} were adopted, and they are indicated as dashed lines in Fig~\ref{fig:Tlock}. If $\tau_\mathrm{lock} \leq 0.1$ Gyr, then the planet is most likely tidally locked (first regime). If $0.1~\mathrm{Gyr} < \tau_\mathrm{lock} < 10~\mathrm{Gyr}$,  the planet may or may not be tidally locked (second regime), and if $\tau_\mathrm{lock} \geq 10$ Gyr, then the timescale is very long and the planet is almost certainly not tidally locked (third regime). In the case with $Q=12$, most of the planets orbiting the inner radius $r_\mathrm{i}$ are in the first regime, with the only exception of V1005 Ori that is in the second regime. At the outer radius $r_\mathrm{o}$, 18 stars are in the first regime, and 12 stars are in the second regime. A threshold can be defined based on the mass, i.e, stars with $M_\star \leq 0.42 M_\odot$ ($M_\star \geq 0.42 M_\odot$) are in the first (second) regime. No stars are in the last regime. In the case with $Q=100$, at $r_\mathrm{i}$, 25 stars are in the first regime, and the rest are in the second. In this case, there is no clear mass threshold that separates both groups, e.g., GJ 846 is in the second regime, and GJ 49 is in the first one, both with $0.57 M_\odot$. No stars are in the third regime. Finally, at $r_\mathrm{o}$, 10 stars are in the first regime, 18 in the second regime, and interestingly, 2 stars in the third regime. There are no clear mass threshold between the groups. The two stars that are most likely not tidally locked are GJ 846 and V1005 Ori, with tidal locking timescales of 12.81 Gyrs and 71.15 Gyrs, respectively. In any case, based on our results, it is likely that the majority of the planets orbiting the HZ ($r_\mathrm{i} \leq r_\mathrm{orb} \leq r_\mathrm{o}$) of one of the stars from our sample will be tidally locked. A similar conclusion was found by \citet{Griessmeir2005} and \citet{Barnes2017}.

\renewcommand{\arraystretch}{1.1} 
\begin{table}[t!]
\centering
\caption{Minimum planetary magnetic field required to sustain a magnetosphere of size $r_\mathrm{M}/r_\mathrm{p}$ at the outer radius of the HZ $r_\mathrm{o}$.}
\begin{tabular}{l|cc}
\hline\hline
GJ, Name & \multicolumn{2}{c}{$B_\mathrm{p,0}^\mathrm{min}(r_\mathrm{o})$} 
\\
\cline{2-3} & 
Case 1 & Case 3 \\
& $r_\mathrm{M}/r_\mathrm{p} = [5,11.7]$ & $r_\mathrm{M}/r_\mathrm{p} = [5,11.7]$\\
\hline
WX UMa & [9.07 , 116.23] & [22.61 , 289.69] \\
GJ 251 & [0.14 , 1.74] & [0.15 , 1.88] \\
GJ 205 & [0.07 , 0.86] & [0.08 , 1.05] \\
DS Leo & [0.26 , 3.27] & [0.15 , 1.95] \\
UV Cet & [10.48 , 134.22] & [43.59 , 558.47] \\
EV Lac & [2.14 , 27.38] & [1.16 , 14.85] \\
GJ 846 & [0.11 , 1.37] & [0.13 , 1.65] \\
V1005 Ori & [0.44 , 5.69] & [0.23 , 2.97] \\
\hline
\end{tabular} 
\tablefoot{The equations are given by $B_\mathrm{p,0}^\mathrm{min}(r_\mathrm{o}) = 2(r_\mathrm{M}/ r_\mathrm{p})^3 B_\star(r_\mathrm{o})$ in Case 1, and $B_\mathrm{p,0}^\mathrm{min}(r_\mathrm{o}) = ((r_\mathrm{M}/ r_\mathrm{p})^6(B_\star(r_\mathrm{o})^2 + 2\mu_0 p_\mathrm{ram}^\mathrm{slow}(r_\mathrm{o}))/f_0^2)^{1/2}$ in Case 3 with slow wind.}
\label{table3}
\end{table}

\subsection{Implications for habitability}

As visible in Table~\ref{table2}, most of our Earth-like hypothetical planets have smaller magnetospheres than that of present-day Earth ($11.7 r_\mathrm{p}$). At $r_\mathrm{orb} = r_\mathrm{i}$ this size is not achieved in any of the analyzed cases. At $r_\mathrm{orb} = r_\mathrm{o}$, this is fulfilled by only one star in Case 1 and in Case 2 with slow wind. No examples of this are found in Case 3. However, it has been suggested that $r_\mathrm{M}/r_\mathrm{p} \approx 5$ is enough to protect the planetary atmosphere \citep[see e.g.,][]{See2014}. This is reasonable, considering that this was also the size of the Earth magnetosphere during the Paleoarchean \citep{Tarduno2010}. In our sample, at $r_\mathrm{orb} = r_\mathrm{i}$, 6 stars achieve this size in Case 1, 13 (11) in Case 2 with the slow (fast) wind, and 11 (9) in Case 3 with the slow (fast) wind. At $r_\mathrm{orb} = r_\mathrm{o}$, this size is achieved by 19 stars in Case 1, 17 (13) in Case 2 with slow (fast) wind, and 16 (13) in Case 3 with slow (fast) wind. In Fig.~\ref{fig:plot5} the magnetopause distances in Case 1 and Case 3 with slow wind are shown as functions of radius of eight stars from our sample. The solid green area represents the HZ $[r_\mathrm{i},r_\mathrm{o}]$. The dotted black lines are related to the regimes of tidal locking, using $Q=100$ (see Section~\ref{TL}), e.g., the zone with “I” represents the first regime of tidal locking. In low mass stars ($M_\star \approx 0.1 M_\odot$) like WX UMa and UV Cet, it is clear that a magnetosphere of $r_\mathrm{M}/r_\mathrm{p} = 11.7$ is quite hard to achieve, especially when the ram pressure of stellar winds is considered. For a planet with $B_\mathrm{p,0} = 1$ G, this can only happen for orbits much further than the extent of the HZ (see Fig.~\ref{fig:plot5}). A potential alternative to create a larger magnetosphere is to increase the planetary magnetic field $B_\mathrm{p,0}$. However, as shown in Table~\ref{table3}, even at the outer edge of the HZ, the planets orbiting these stars would require very strong magnetic fields, at least two orders of magnitude stronger than those considered in our estimations. Another issue is that in these cases, a planet in the HZ would be most likely tidally locked (first regime). The cases of GJ 251, EV Lac, and DS Leo are somewhat more favorable. In EV Lac, a planet with $r_\mathrm{M}/r_\mathrm{p} = 5$ would still be close to the outer edge of the HZ, especially in Case 3. Furthermore, the planet would be in the second regime of tidal locking, making habitability somewhat more favorable. In GJ 251 and DS Leo the situation is similar; a planet can still potentially be in some part of the HZ with a magnetosphere between $5 r_\mathrm{p}$ and $11.7 r_\mathrm{p}$, in the second regime of tidal locking.

Finally, GJ 205, GJ 846 and V1005 Ori provide most likely the best scenarios. A planet in the HZ of GJ 205 can have a magnetosphere with the size of modern Earth, and be in the second tidal locking regime. GJ 846 and V1005 Ori have a region of their HZ in the third tidal locking regime. Moreover, a planet orbiting in this area will have a magnetosphere larger than $5 r_\mathrm{p}$ in Cases 1 and 3. From Table~\ref{table3} it is also visible that for these stars, the magnetic field required for a planet to achieve $r_\mathrm{M}/r_\mathrm{p} = 11.7$ orbiting the outer radius $r_\mathrm{o}$, is certainly not as strong as in the low mass cases previously mentioned.

\section{Discussion\label{discussions}}
In our model, we neglected the ram pressure due to atmospheric escape of the exoplanet. In the case of Earth, this term can be safely neglected, because the atmospheric escape rate on Earth is relatively low ($\sim 1.4$ kg/s; \citealt{Gunell2018, 2025RvMPP...9...18H}). However, in the context of exoplanetary systems, this might not always be the case as the atmospheric mass loss can be orders of magnitude higher than Earth’s  \citep{2007AsBio...7..167K,2018PNAS..115..260D}. The assumption $B_\mathrm{p}^2/2\mu_0 \gg p_\mathrm{p}^\mathrm{ram}$ is valid when $ \dot{M}_\mathrm{p}^\mathrm{crit} \ll 2 \pi B_\mathrm{p}^2 r^2 / \mu_0 v_\mathrm{p}$. To estimate such critical value, we assume $B_\mathrm{p} = 1$ G, $r=5 r_\mathrm{p} = 5 R_\oplus$, and an outflow speed of $v_\mathrm{p} \sim 10$ km/s \citep[e.g.][]{2019A&A...624L..10J}. These values yield $\dot{M}_\mathrm{p}^\mathrm{crit} \ll 8 \cdot 10^4~\mathrm{kg/s}$, which is much larger than Earth's value. However, we note that in strongly irradiated planets, $\dot{M}_\mathrm{p}$ could be larger than $\dot{M}_\mathrm{p}^\mathrm{crit}$ \citep[e.g.][]{2018PNAS..115..260D,2019A&A...624L..10J}. In such cases, our assumption is no longer valid, as the planetary outflows would affect the magnetopause and the corresponding ram term should be included. We postpone this regime for future studies.

In Section~\ref{results}, we showed that in some cases, mainly low mass stars, the ram pressure of stellar winds can affect the size of the magnetosphere significantly. However, it should be noted that there is a lot of uncertainty in the characterization of stellar winds in M dwarfs, therefore, it is worth exploring parameters obtained with different methods to check how the results can potentially change. Simulations of \cite{Cohen2014} found $v_\mathrm{wind}= 300~\mathrm{km/s}$ and $\dot{M} = 3 \cdot 10^{-14}~M_\odot\,\mathrm{yr^{-1}}$ for EV Lac, which yields $r_\mathrm{M}/r_\mathrm{p}(r_\mathrm{i},r_\mathrm{o}) = (5.688,7.099)$ in Case 2, and $r_\mathrm{M}/r_\mathrm{p}(r_\mathrm{i},r_\mathrm{o}) = (3.278,5.025)$ in Case 3, being slightly bigger than our estimations.  \cite{Wood2021} published observational constrains of the winds from four M dwarfs from our sample, i.e., EV Lac, GJ 205, YZ CMi and AD Leo. Unfortunately, the mass loss rate of AD Leo could not be determined, and therefore, we will not include it in the analysis. Using the provided values for $v_\mathrm{wind}$ and $\dot{M}_\star$, it yields $r_\mathrm{M}/r_\mathrm{p}(r_\mathrm{i},r_\mathrm{o}) = (8.349,10.421)$, $(11.623,14.463)$, $(5.178,6.482)$ in Case 2, and $r_\mathrm{M}/r_\mathrm{p}(r_\mathrm{i},r_\mathrm{o}) = (3.296,5.126)$, $(9.769,13.526)$, $(3.010,4.636)$ in Case 3, for EV Lac, GJ 205 and YZ CMi, respectively. All these values are larger than our original estimations. In EV Lac, Case 2, our original estimation is only $54 \%$ of the value provided above, being the largest difference in both cases. As mentioned in Section~\ref{ram}, we neglected the thermal pressure. If we include it in our formulation, defining $p_\star^\mathrm{th}(r) = 2n(r) k_\mathrm{B}T(r)$, where $n(r)$ is the number density, $k_\mathrm{B}$ is the Boltzmann constant and $T(r)$ is the temperature profile, obtained following \cite{Johnstone2015a}, we find that $p_\mathrm{tm}$ typically corresponds to less than 2\% of the stellar ram pressure. This translates to new magnetopause distances corresponding to 99.61-99.96\% of their original value in Case 2, and ~99.78-99.99\% in Case 3, both at $r_i$ and with slow wind.

Using spherical coordinates, the (isotropic) stellar wind velocity can be written as $\bm{v}_\mathrm{wind}(r) = v_r(r) \bm{\hat{r}} + v_\phi(r) \bm{\hat{\phi}}$. In Cases 2 and 3 we assumed that the wind velocity has only a radial component, where $v_\mathrm{wind} \approx v_r = \alpha v_\mathrm{esc}$, neglecting the azimuthal component. However, as the hypothetical planets are orbiting very close to its host star, the relative velocity between the wind and the planet includes not only the radial wind speed but also the orbital (Keplerian) velocity of the planet, since from the planet’s frame the wind acquires a tangential component. Assuming $v_\phi \approx v_\mathrm{orb} = \sqrt{GM_\star/r_\mathrm{orb}}$ our results do not change significantly, and the magnetospheres are typically $\sim 99.8\%$ of their original value. Alternatively, if we define $v_\phi = \Omega_\star r_\mathrm{orb}$ \citep{1958ApJ...128..664P, 2019winds}, our results change typically between $97-99\%$. The largest differences are coming from YY Gem A and YY Gem B, where the magnetospheres are 90\% of their original value. It should be noted that to use $v_\phi = \Omega_\star r_\mathrm{orb}$ we assumed that $r_\mathrm{orb}$ is always smaller than the Alfvén radius, if this is not the case, then the component decays with the inverse of the distance \citep[see e.g.,][]{2019winds}. Additionally, we also assumed that the wind reached its terminal velocity, and therefore, it is constant with distance. In solar-like stars this is a reasonable typical assumption. However, in the context of M dwarfs this might not always be valid, as the wind could still be accelerating when it reaches the hypothetical planet. One approach to this issue, is to define a radial profile for the wind velocity \citep[e.g.,][]{Vidotto2014, Johnstone2015a}. In any case, we do not expected our result to change significantly after the inclusion of such profile. 

In Section~\ref{TL}, we conclude that most of the planets orbiting the HZs of our stellar sample might be tidally locked. This might have some implications in the magnetosphere that we did not considered in our estimations. For example, we assumed an Earth-like magnetic field, and it is well known that the geomagnetic field is generated by a self-sustained dynamo hosted in the outer liquid iron core of Earth \citep{1978mfge.book.....M}. This dynamo can be strongly affected by Earth’s rotation \citep[see e.g.,][]{2011PEPI..187..157C}. In our models, after tidal locking, the synchronous rotation rate of the planet is given by Eq.~(\ref{omegaf}), and it should be noted that $\omega_\mathrm{f}$ can be significantly lower than the initial $\omega_\mathrm{i}$ considered. For example, at $r_\mathrm{orb} = r_\mathrm{i}$ in low mass stars ($M_\star \leq 0.14 M_\odot$), the ratio $\omega_\mathrm{i}/\omega_\mathrm{f}$ is typically 5-10, while in stars with $M_\star > 0.42 M_\odot$, this ratio can span 2 orders of magnitude. To quantify how much rotation affects the geodynamo, some scaling laws can be used \citep[see][]{2010SSRv..152..565C}. For example, using the scaling proposed by \cite{1976PEPI...12..350B}, i.e., $B_\mathrm{p}^2 \propto \rho_\mathrm{c} \omega_\mathrm{p}^2 R_\mathrm{c}^2$, where $\omega_\mathrm{p}$ is the rotation rate of the planet, $\rho_\mathrm{c}$ is the density of the core and $R_\mathrm{c}$ is its radius, we find that in the worst case ($\omega_\mathrm{i} = 100\omega_\mathrm{f}$) the magnetosphere radius shrinks by a factor of 4.64, and 1.70 in a generous case ($\omega_\mathrm{i} = 5\omega_\mathrm{f}$). Therefore, it should be noted that planets with an orbit in the first or second regime of tidal locking might have a smaller magnetosphere than our initial estimations, especially if the host star has $M_\star > 0.42 M_\odot$. In any case, quantifying the impact of tidal locking, strongly depends on the chosen scaling law. For example, the relation proposed by \cite{2006E&PSL.250..561O} suggests that $B_\mathrm{p}$ depends mainly on the convective energy flux available in the core, and rotation might play a role only in the geometry of it. Assessing the impact of tidal locking if different “rotation based” and “energy-flux based” scaling laws are considered is postponed to future studies.

\section{Summary and conclusions \label{conclusions}}
We estimated the magnetopause distances of an Earth-like planet orbiting the inner $r_\mathrm{i}$ and outer $r_\mathrm{o}$ radii of the HZ of a sample of 30 M dwarfs. In the case where only the magnetic pressure of the stellar wind is considered (Case 1), we find that $63.33\%$ of the hypothetical planets orbiting $r_\mathrm{o}$ have magnetospheres larger than $5 r_\mathrm{p}$, which is believed to be enough to protect the atmosphere of the planet \citep{Tarduno2010,See2014}. If only the ram pressure of the stellar wind is considered (Case 2), this size is achieved by $56.66\%$ ($43.33\%$) of our sample assuming slow (fast) winds. In the case where both pressures are considered (Case 3), these percentages stay almost the same, i.e., $53.33\%$ ($43.33\%$) assuming slow (fast) winds . Our estimations in Case 1 typically agree with the results of \cite{Vidotto2013} with a few exceptions, like DS Leo and DT Vir. In low mass stars ($M < 0.15 M_\odot$), the ram pressure of the stellar winds seems to affect the size of the magnetosphere more than their magnetic pressure (see Fig.~\ref{fig:plot2}). In some cases, the inclusion of dynamical pressure can be somewhat beneficial for the magnetosphere. This is a consequence of the form factor included in our estimations, representing the contribution of the Chapman-Ferraro Currents to the planetary magnetic field. This is better visible in the estimations with slow wind. In the cases with fast wind, the ram pressure increases, reducing the size of the magnetopause, despite the shielding effect produced by the magnetopause currents. At the inner radius $r_\mathrm{i}$ the situation is typically less favorable, and only $20\%$ or our sample have $r_\mathrm{M}/r_\mathrm{p} \geq 5$ in Case 1, $43.33\%$ ($36.66 \%$) in Case 2 with slow (fast) wind, and $36.66\%$ ($30 \%$) in Case 3 with slow (fast) wind. At $r_\mathrm{o}$ only GJ 205 can host a planet with a magnetosphere comparable to that of modern Earth ($r_\mathrm{M}/r_\mathrm{p} \geq 11.7$). This is not achieved in any case at $r_\mathrm{i}$. 

We conclude that deviations from the Parker spiral can affect the results significantly. In Case 1, the magnetosphere gets reduced by $30-40 \%$ at $r_\mathrm{i}$, and by $40-45 \%$ at $r_\mathrm{o}$, leading to no planets achieving $r_\mathrm{M}/r_\mathrm{p} \geq 5$ at $r_\mathrm{i}$, and only $13.33 \%$ achieving this at $r_\mathrm{i}$. In Case 3, at $r_\mathrm{o}$, this condition is fulfilled by $23.33\%$ ($13.33\%$) of the planets, considering slow (fast) wind. At $r_\mathrm{i}$ this is achieved only by four of them in Case 3. In reality, these deviations can increase other parameters, like the wind velocity or mass loss rate, decreasing the magnetospheres more than what we estimated. Additionally, we conclude that majority of planets orbiting the HZ of our sample might be tidally locked. Only two stars have the outer radius of the HZ in a region where a planet will be most likely not tidally locked (GJ 846 and V1005 Ori). In any case, tidal locking does not necessarily imply that the planet is not habitable. For example, circulation in the planetary boundary layer and in the oceans can distribute the heat from the dayside to the nightside, preventing the atmospheric collapse  \citep{2014PNAS..111..629H, 2015ApJ...806..180W}

In real systems, many additional factors can affect the magnetosphere. While our estimations provide a basic assessment of the effects of stellar magnetic and ram pressures, a realistic formulation of the problem should include bow shocks \citep[e.g.][]{2012JGRA..117.5208J}, and more complex current systems that were not included in this work \citep[see e.g., Fig.~1 of][]{2017SSRv..206..521L}. Furthermore, as recently reported by \cite{Ilin2025}, planets orbiting close enough to their host star can enhance its stellar activity, and induce flares. CMEs and strong flares are also thought to affect the magnetosphere \citep{2007AsBio...7..167K, 2007AsBio...7..185L}. However, it is unclear if they are a major threat to habitability, as it has been suggested that CMEs from M dwarfs are much less powerful than the predictions based on the solar analogy \citep[e.g.][]{Wood2021}. In any case, CMEs in M dwarfs are still under debate and are highly speculative.

The star-planet interaction can also affect close-in planets and their atmospheres significantly. For example, it can lead to substantial heating via ohmic dissipation, depending on the resistivity of the planet \citep{2012ApJ...745....2L}. For a dipolar planetary magnetic field, like in our estimations, the topology of the interaction plays a critical role, as the inclination of the dipolar field determines the location of the reconnection sites between the stellar and planetary fields \citep{2015JGRA..120.5645L, 2018haex.bookE..25S}. The strength of the planetary magnetic field also plays an important role in our estimations (see Table~\ref{table3}). Additionally, it can affect the geometry of the atmospheric escape of the planet \citep{2015ApJ...813...50K, 2021MNRAS.508.6001C}. Currently, detecting the magnetic field of exoplanets is still very challenging, and while there are a few promising methods, no unambiguous measurement has been done \citep{2024arXiv240415429B, strugarek2025}. Future observations will be useful to constrain the range of planetary magnetic fields and get more solid estimations of the parameters that characterize the magnetospheres of exoplanets.

\begin{acknowledgements}
We thank the anonymous referee for providing useful and constructive comments on the manuscript. JPH acknowledges financial support from ANID/DOCTORADO BECAS CHILE 72240057. DRGS gratefully acknowledges support by the ANID BASAL project FB21003 and via the  Alexander von Humboldt - Foundation, Bonn, Germany.
\end{acknowledgements}

\bibliographystyle{aa}
\bibliography{paper.bib}

\end{document}